\documentclass[prl,twocolumn,superscriptaddress,showpacs,floatfix]{revtex4}
\usepackage{graphicx}
\begin{document}

\title{Quantum and Classical Glass Transitions in $\text{LiHo}_x\text{Y}_{1-x}\text{F}_4$}
\date{\today}
\author{C. Ancona-Torres}
\author{D.M. Silevitch}
\affiliation{The James Franck Institute and Department of Physics, The University of Chicago, 
Chicago, IL 60637}
\author{G. Aeppli}
\affiliation{London Centre for Nanotechnology and Department of Physics and Astronomy, UCL, 
London, WC1E 6BT, UK}
\author{T.F. Rosenbaum}
\email{tfr@uchicago.edu}
\affiliation{The James Franck Institute and Department of Physics, The University of Chicago, 
Chicago, IL 60637}

\begin{abstract}
When performed in the proper low field, low frequency limits, measurements of the dynamics and the nonlinear susceptibility in the model Ising magnet in transverse field, $\text{LiHo}_x\text{Y}_{1-x}\text{F}_4$, prove the existence of a spin glass transition for $x$ = 0.167 and 0.198. The classical behavior tracks for the two concentrations, but the behavior in the quantum regime at large transverse fields differs because of the competing effects of quantum entanglement and random fields. 
\end{abstract}
\pacs{73.43.Nq, 75.10.Nr, 75.50.Dd, 75.50.Lk}

\maketitle

Research on spin glasses \cite{Mydosh72} has led not only to deep insights into disordered materials and the glassy state, but has generated novel approaches to problems ranging from computer architecture through protein folding to economics. The rugged free energy landscape characteristic of such systems defies usual equilibrium analyses, with pronounced non-linear responses and history dependence. At low temperatures, and in cases where barriers to relaxation are tall and narrow, quantum mechanics can enhance the ability to traverse the free energy surface \cite{Ray89}. The $\text{LiHo}_x\text{Y}_{1-x}\text{F}_4$ family of materials represents the simplest quantum spin model, the Ising magnet in transverse field, and it has been an especially useful system to probe the interplay of disorder, glassiness, random magnetic fields and quantum entanglement \cite{Wu91,Wu93,Reich87,Ghosh03,Schechter06,Gingras06,Biltmo07,Bhatt85}. The parent compound, $\text{LiHoF}_4$, is a dipole-coupled Ising ferromagnet with Curie temperature $T_C = 1.53$~K. Applying a magnetic field $H_t$ transverse to the Ising axis introduces quantum mixing of classical spin-up and spin-down eigenstates which are split by an energy $\Gamma\sim H_t^2$, or equivalently, tunes the tunneling probabilities for walls between patches of ordered spins \cite{Brooke01}. Hence, quantum fluctuations controllable by an external field can drive the classical order-disorder transition to zero temperature, resulting in a much studied ferromagnetic quantum critical point \cite{Bitko96,Ronnow05,Chakraborty04}. 

\begin{figure}
\includegraphics[scale=0.45,clip=true,bb=20 80 600 500]{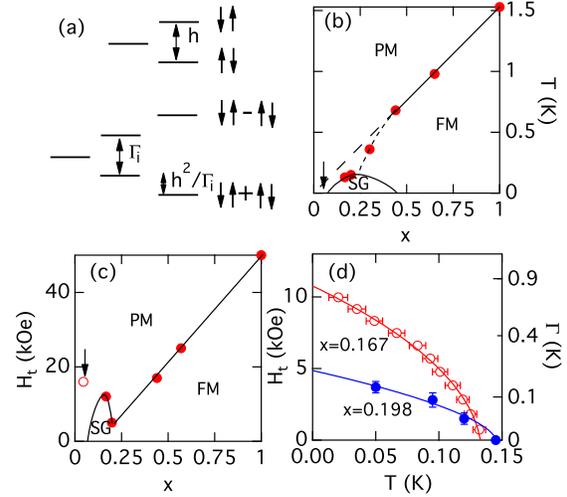}
\caption{\label{fig:phase}(color online) Energy levels and phase diagram for $\text{LiHo}_x\text{Y}_{1-x}\text{F}_4$ (a) Schematic of energy levels for antiferromagnetically coupled spins in a (top) random field and (bottom) uniform transverse field $\Gamma$, and then a random field $h$. (b) Magnetic phases in the $x-T$ plane. Arrow denotes spin liquid ``anti-glass'' phase. (c) Magnetic phases in the $x-H_t$ plane. Open circle shows peak in $\chi^\prime(H_t)$ for the antiglass \cite{Silevitch07a}. (d) Spin glass/paramagnet phase boundaries for $x=0.167$ and 0.198 cross.}
\end{figure}

The nature of the ground state can be tuned by partially substituting non-magnetic Y for the magnetic Ho \cite{Reich90}. Dilution enhances the effects of the frustration inherent in the dipolar interaction, with the ferromagnet giving way to a spin glass at $x\sim 0.2$ (Fig.~\ref{fig:phase}). The transverse component of the dipolar coupling introduces other phenomena. At large x, the ground state becomes a disordered ferromagnet, with the low temperature dynamics dominated by domain wall tunneling \cite{Brooke01}. The high $T$ and low $H_t$ behavior reveal both the effects of Griffiths singularities \cite{Guo96} and the internal random fields due to the application of a uniform $H_t$ to a disordered Ising magnet \cite{Schechter06,Gingras06,Silevitch07}. For $x \stackrel{<}{\sim} 0.1$, the internal transverse fields, $\Gamma_i$, induce quantum entanglement that prevents the system from freezing and stabilizes a spin liquid ``antiglass'' phase down to very low temperatures \cite{Reich87,Ghosh03,Silevitch07a}. In the intermediate range, where the spin glass phase is stable, the tendencies towards ferromagnetism and random field effects found at high x  compete with the massive quantum entanglement of the antiglass.  This competition might be expected to lead to very different statics and dynamics for small changes in x. Fig. \ref{fig:phase}a illustrates the essential physics. In the classical ($\Gamma_i=0$) case, a small ``random'' field of strength $h$ at site 1 and 0 at site 2 will produce a splitting of order $h$ between the degenerate $|\uparrow\downarrow\rangle$, $|\downarrow\uparrow\rangle$  classical ground states. On the other hand, $H_t\neq0$ will yield a single non-degenerate ground state $|\uparrow\downarrow\rangle + |\downarrow\uparrow\rangle$,  on which the only effect of a small $h$ will be an energy change $\sim h^2$.

The first work on the spin glass state in $\text{LiHo}_{0.167}\text{Y}_{0.833}\text{F}_4$ revealed a well-defined transition from the paramagnet to the glass, evidenced by both a sharp divergence in the nonlinear susceptibility, $\chi_3$, and a dynamical signature in the dissipative component of the linear susceptibility, $\chi^{\prime\prime}$ \cite{Wu91,Wu93}. These measurements relied on $H_t$ to speed up the system dynamics to the point where the $f\rightarrow0$ limit could be probed directly, and led to considerable theoretical work \cite{Bhatt85,Guo96,Ye93}, some with good qualitative agreement with our experiments \cite{Cugliandolo01}. Within the last three years, the concept that randomly placed classical dipoles \cite{Stephen81} should undergo a spin glass transition has itself been questioned on account of numerical work on small cubic lattices \cite{Snider05}. More recent experiments \cite{Jonsson07} employing a $\mu$SQUID magnetometer did not reveal a divergence in $\chi_3$, leading the authors to a similar conclusion, namely that $\text{LiHo}_{0.167}\text{Y}_{0.833}\text{F}_4$ is not a spin glass. Unfortunately, the authors of \cite{Jonsson07} used large longitudinal fields and fast sweep rates, probing the system very far from equilibrium, and thus obscured the meaning of their data. In this Letter, we show explicitly that when data are acquired in the proper small longitudinal field, low frequency limit, clear evidence is seen for a spin glass transition. Moreover, we report the discovery that a minor change in $x$ from 0.167 to 0.198 - approaching the multicritical point where spin glass and ferromagnetic phases coexist (Fig.~\ref{fig:phase}b) - results in dramatic changes in the quantum ($H_t$-dependent) behavior. 

We performed ac susceptibility measurements from 1 to $10^5$ Hz on single crystal needles of $\text{LiHo}_x\text{Y}_{1-x}\text{F}_4$ mounted on the cold finger of a dilution refrigerator. The magnitude of $\chi_1$ was consistent with the Curie-Weiss law for $\text{Ho}^{3+}$ ions at high $T$. The susceptibility was corrected using the demagnetization factor of rods of the same aspect ratio. Ho concentrations $x$ were determined to $\pm0.001$ by a differential weighing technique. Static transverse magnetic fields up to 80 kOe and longitudinal fields, $h_\ell$, up to 300 Oe were supplied by a superconducting solenoid and Helmholtz coils, respectively. The ac excitation amplitude was restricted to less than $A = 0.02$~Oe to ensure linear response and to control heating. For measurements of the nonlinear susceptibility, the dc longitudinal field was swept at 0.04 Oe/s so that this rate was smaller than the effective sweep rate at $f = 1.5$~Hz of $2\pi Af = 0.2$~Oe/s.

We plot in Fig.~\ref{fig:phase}d the $T$-$H_t$ phase diagrams for both the x = 0.167 and 0.198 spin glasses. The transition is defined by the emergence of a flat spectral response at low $f$ in $\chi_1^{\prime\prime}$, corresponding to $1/f$ noise in the magnetization. This dynamically-determined phase boundary coincides with that derived from the maxima of $\chi_3(f\rightarrow0)$. The classical spin glass transition $T_g(H_t=0)$ increases with increasing $x$, but lies below the mean-field ferromagnetic $T_C(x)=xT_C(x=1)$. However, once the transverse magnetic field is turned on and the relative importance of quantum entanglement and random field effects becomes germane, the samples respond very differently and the phase boundaries actually cross. 

\begin{figure}
\includegraphics[scale=0.42]{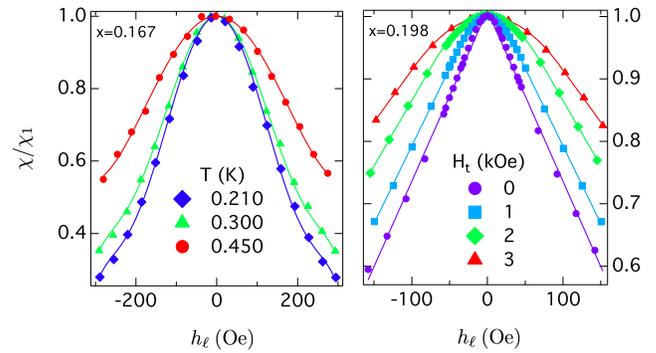}
\caption{\label{fig:parab}(color online) Scaled susceptibility for $x = 0.167$, $H_t=2.1$~kOe (a) and $x = 0.198$, $T=0.25$~K (b) as a function of longitudinal bias field added to sinusoidal 1.5 Hz probe field with amplitude of 0.02 Oe. Both exhibit pronounced non-linear responses, but with different forms most likely due to random field effects \cite{Silevitch07} appearing for the more concentrated sample. }
\end{figure}

Fig.~\ref{fig:parab} illustrates the pronounced sensitivity to fields applied parallel to the Ising axis and the evolution of the nonlinear response with x. $\chi(h_\ell)$ for x = 0.167 can be described by a conventional power series expansion:  $\chi  = \chi _1  - 3\chi _3 h_\ell ^2  + 5\chi _5 h_\ell ^4  +  \cdots$, with all orders of the susceptibility growing as the glass transition is approached from above. The longitudinal field dependence of the susceptibility for x = 0.198 also exhibits strong nonlinearities, but has qualitatively different behavior. The parabolic $\chi_3$  at small $h_\ell$  rolls over to a linear dependence at large field, consistent with a tendency towards the singular linear behavior, attributed to random fields, seen for the disordered ferromagnet with x = 0.44 in the classical low $H_t$, high $T$ regime \cite{Silevitch07}. The dramatic change in $\chi_3$, paired with the phase boundary crossing in Fig.~\ref{fig:phase}d, represents the major new discovery of our present work, and points to significant changes in the underlying physical mechanisms arising from a small change in x near the onset of ferromagnetic long-range order.

\begin{figure}
\includegraphics[scale=0.45]{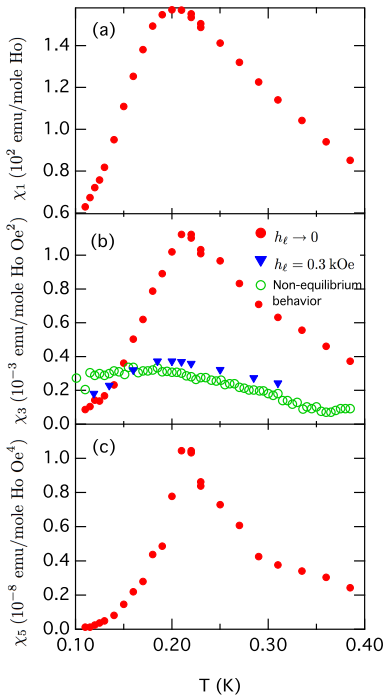}
\caption{\label{fig:nonlinear}(color online) Linear and nonlinear susceptibilities for $\text{LiHo}_{0.167}\text{Y}_{0.833}\text{F}_4$ derived from data akin to Fig.~\ref{fig:parab}a and fit to $\chi  = \chi _1  - 3\chi _3 h_\ell ^2  + 5\chi _5 h_\ell ^4  +  \cdots$. In (b) we demonstrate the danger of not explicitly accounting for high order, non-linear effects. Fitting our data to a simple parabola out to 0.3 kOe or rapidly sweeping the field as in Ref.  \cite{Jonsson07} (open circles) suppress the divergence of $\chi_3$ and mask the true physics.}
\end{figure}

\begin{figure}
\includegraphics[scale=0.45]{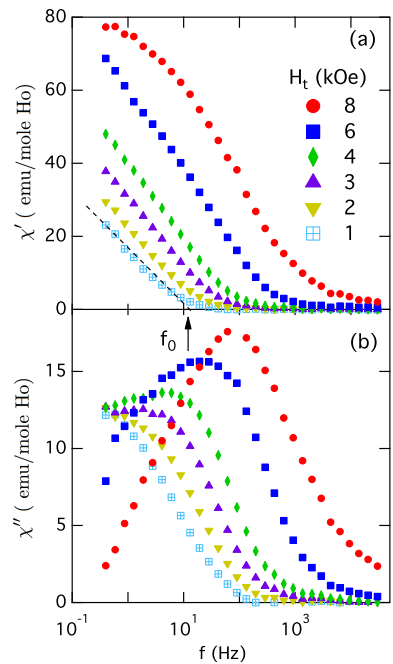}
\caption{\label{fig:freq}(color online) Spectral response of the magnetic susceptibility of $\text{LiHo}_{0.198}\text{Y}_{0.802}\text{F}_4$ at $T=0.05$~K for a series of transverse fields. The spin glass transition is marked by a flat low-frequency response in $\chi_1^{\prime\prime}$, corresponding to a logarithmic dependence of $\chi_1^\prime$ with onset $f_0$ and, via the fluctuation-dissipation theorem, $1/f$ noise in the magnetization.  }
\end{figure}

We show in Fig.~\ref{fig:nonlinear} the temperature dependence of the linear and nonlinear terms in the susceptibility for $x = 0.167$ at $H_t = 2.1$~kOe. As expected, increasing orders of the susceptibility diverge increasingly more strongly \cite{Levy86}, reflecting the approach to a phase transition. This is in accord with the results reported by Wu et al. \cite{Wu93} and in disagreement with the recent results of J\"onsson et al. \cite{Jonsson07}. The discrepancy can be understood by looking at the different limits in which the system was examined. In Ref. \cite{Wu93} and in the present work, great care was taken to accumulate data in the $h_\ell\rightarrow0$  limit ($\pm10$ - $20$ Oe about the peak). By contrast, the $\mu$SQUID technique of Ref. \cite{Jonsson07} involved polarizing the system in a large longitudinal field (3 kOe, 15 times the scale of Fig.~\ref{fig:parab}), and then rapidly decreasing the field through zero at sweep rates of up to 50 Oe/s, corresponding to effective frequencies in this work of over 100 Hz. As can be seen from Fig.~\ref{fig:freq}, frequencies in this range are far from the equilibrium limit even in the presence of substantial transverse fields, suggesting that the results reported in \cite{Jonsson07} do not capture the true physics.

The differences in the underlying mechanisms that were observed in the nonlinear susceptibilities also emerge in dynamical measurements of the two concentrations. These can be determined by examining the spectroscopic response of the system over several decades of $f$, as shown in Fig.~\ref{fig:freq} for $x=0.198$ at a series of $H_t$. This sample is slower than its $x = 0.167$ counterpart \cite{Wu91}, where quantum fluctuations promoted by the off-diagonal elements of the dipolar interaction more effectively speed the long-time relaxation. We characterize the approach to the spin glass from above by fitting the low frequency tail of $\chi_1^{\prime\prime}$ to a power law form, $f^\alpha$. $T_g(H_t)$ and $H_c(T)$ are defined dynamically when $\alpha\rightarrow0$  and fluctuations occur on all (long) timescales. At this point $\chi_1^\prime$ grows logarithmically with $f$, where the onset frequency $f_0$ defines the fastest relaxation process available to the system and characterizes the quantum tunneling rate. 

\begin{figure}
\includegraphics[scale=0.4]{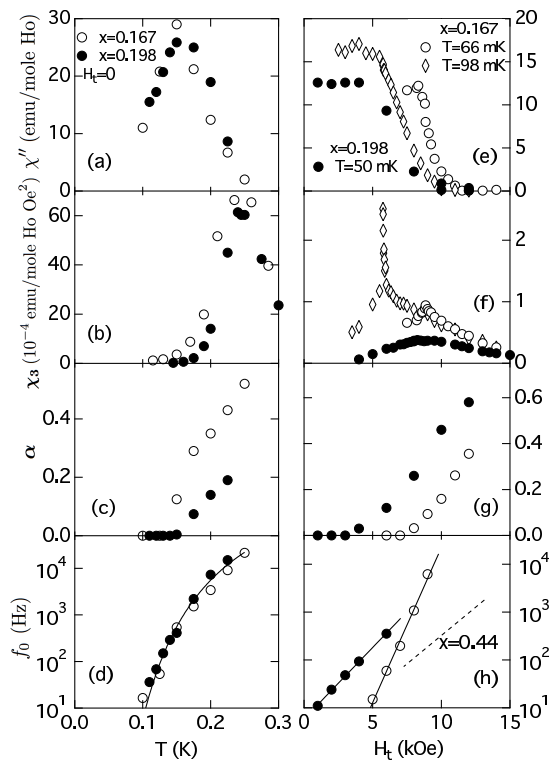}
\caption{\label{fig:summary}Dynamical and non-linear signatures of the spin glass transition in the classical ($H_t=0$, a--d) and quantum ($T\rightarrow0$, e--h) regimes. The contrasting behavior of the two concentrations with transverse field reflect the competing effects of quantum entanglement at small $x$ and random fields at large $x$ (see text). x=0.167 data in e and f was previously published in  \cite{Wu93}. Onset of non-zero $\alpha$ in g corresponds to $H_c$. Solid lines in d and h are Arrhenius and WKB fits, respectively. Dashed line is the slope for $x=0.44$ \cite{Brooke01}.}
\end{figure}

The quantities derived from the nonlinear measurements and the spectroscopic response are combined in Fig.~\ref{fig:summary}. Magnetic glass transitions for both x = 0.167 and 0.198 are defined classically (a-d) and quantum-mechanically (e-h) by $\alpha\rightarrow0$ (c,g). When it is possible to reach the $f\rightarrow0$ limit at modest $H_t$, then a sharp, dynamical feature in $\chi_1^{\prime\prime}$ (a,e) and a divergence of $\chi_3$  (b,f) also serve to define $T_g(H_t)$. Both concentrations show similar behavior in the classical limit, but differ substantially in the quantum limit, underscoring the crossover in the underlying physics. The onset frequency for relaxation, $f_0$, follows an Arrhenius law, $e^{-\Delta/k_BT}$, in the classical limit where thermal fluctuations dominate (d) and a WKB form in the quantum limit (h) \cite{Brooke01}. As a function of $T$ with $H_t=0$, $f_0$ for the two concentrations are indistinguishable. By contrast, at base temperature the $H_t$-dependent $f_0$ curves have substantially different slopes for the two values of $x$, with faster relaxation at low $H_t$ in the x = 0.198 sample with the suppressed quantum glass transition. The slope seen for x = 0.198 is similar to what was previously observed for the x = 0.44 ferromagnet \cite{Brooke01}, suggesting that the random-field effects which play an important role in the dynamics of the ferromagnet are also significant in the x = 0.198 glass.

We have verified that in low transverse fields the dilute dipolar-coupled magnet, $\text{LiHo}_x\text{Y}_{1-x}\text{F}_4$, can display the static and dynamic signatures of a conventional spin glass. These results are in agreement both with early theory \cite{Stephen81} and experiments of fifteen years ago, and in disagreement with incorrect (as shown above) conclusions drawn from interesting recent experiments \cite{Jonsson07} where the same material was subjected to very strong and rapid perturbations away from equilibrium. While the static signature of the spin glass transition - a diverging non-linear susceptibility - seems to disappear for high $H_t$, the dynamical signature - the appearance of a flat  $\chi^{\prime\prime}(f\rightarrow0)$ - persists and indeed becomes sharper. This suggests that internal random fields \cite{Schechter06,Gingras06,Silevitch07} notwithstanding, there is a distinct quantum glass state that can be entered via a first order transition \cite{Wu91,Cugliandolo01} for which $\chi_3$  would not diverge. The striking new discovery that we make here is that the (quantum) critical field $H_c(T\rightarrow0)$ for this state is a non-monotonic function of $x$, with a lower value for $x = 0.198$ than for both the $x = 0.167$ spin glass and the $x = 0.44$ ferromagnet. We suspect that for $x = 0.198$ the random field effects seen near the Curie point for the $x = 0.44$ sample are important, and suppress the magnetic glass phase. On the other hand, the glass-like state for $x = 0.167$ is more robust because of the larger quantum entanglement derived from the relatively greater population of antiferromagnetically coupled spins; these entanglement effects are known to play a major role in the dynamics of the 0.045 spin liquid \cite{Ghosh03}. As $x$ is lowered even more, we land in the antiglass spin liquid phase. The quantum spin glass then acquires new meaning as a valence bond glass \cite{Bhatt97} of a type where there are multiple ways of drawing the bonds to construct pairs, while the spin liquid for $x = 0.045$ is a non-degenerate liquid with a unique pattern of valence bonds.

The work at the University of Chicago was supported by NSF MRSEC Grant No. DMR-0213745 and that at UCL by EPSRC Grant No. EP/D049717/1. D.M.S. acknowledges support from DOE BES Grant No. DE-FG02-99ER45789.

\end{document}